\documentclass[pre,twocolumn,nopacs,amssymb]{revtex4}

\usepackage{graphicx}
\usepackage{dcolumn}
\usepackage{bm}

\def\gs{\gtrsim}
\def\ls{\lesssim}
\def\be{\begin{equation}}
\def\en{\end{equation}}                  
\def\p{\partial} 
\newcommand{\bi}[1]{\mbox{\boldmath$#1$}}
\newcommand{\av}[1]{\langle{#1}\rangle}

\def\bea{\begin{equation}\begin{array}{rcl}}
\def\ena{\end{array}\end{equation}}

\def\q{{\footnotesize{\it q}}\kern -5pt {\footnotesize{\it q}}}
\def\k{{\footnotesize{\it k}}\kern -5pt {\footnotesize{\it k}}}
\def\seq{\sim \kern -12pt \lower 5pt \hbox{$\displaystyle =$}}
\def\nnabla{\nabla\kern-3.3mm\nabla}
\def\ge{> \kern -12pt \lower 5pt \hbox{$\displaystyle =$}}
\def\le{< \kern -12pt \lower 5pt \hbox{$\displaystyle =$}}
\def\gs{> \kern -12pt \lower 5pt \hbox{$\displaystyle{\sim}$}}
\def\ls{< \kern -12pt \lower 5pt \hbox{$\displaystyle{\sim}$}}

\def\be{\begin{equation}}
\def\bea{\begin{eqnarray}}
\def\en{\end{equation}}
\def\ena{\end{eqnarray}}
\def\p{\partial }
 
\renewcommand{\theequation}{\arabic{section}.\arabic{equation}}
\renewcommand{\bf}{\bm}

\begin{document}
\bibliographystyle{prsty}

\title{Intermediate states  at structural 
phase transition: Model with a one-component order parameter 
coupled to  strains}
\author{Akihiko Minami  and Akira Onuki}
\affiliation{Department of Physics, 
Kyoto University, Kyoto 606-8502, Japan}
\date{\today} 

\begin{abstract}
We study a Ginzburg-Landau model of structural phase transition 
 in two dimensions, 
in which a single   order parameter is coupled to the tetragonal and 
dilational strains. Such elastic coupling terms in the free energy 
much affect the phase transition behavior 
particularly near  the tricriticality. 
A characteristic feature  is 
 appearance  of intermediate states,  
where   the ordered and  disordered regions 
 coexist on mesoscopic  scales 
in nearly steady states in a temperature window. 
The window  width increases with increasing the strength 
of the dilational coupling. 
It arises from freezing  of  phase ordering  in  
inhomogeneous   strains.  No 
impurity mechanism is involved.
 We present a simple theory 
of the intermediate states to produce phase diagrams consistent 
with simulation  results.
\end{abstract}

\pacs{}

\maketitle
\setcounter{equation}{0}

\section{Introduction}

At structural phase transitions, 
 anisotropically deformed 
domains of the low-temperature ordered phase emerge in the 
high-temperature disordered  phase, as the temperature is 
lowered  \cite{Kha,Onukibook}.  In these processes,  
elastic strains are induced around the domains,  
radically influencing  the phase transition behavior. 
 The order parameter 
can be the concentration, the atomic configuration 
in a unit cell, the electric polarization, the magnetization, 
 the orbital order, etc.   
Thus the effects are  extremely varied and   
complex  in real solids and  have  not 
yet been well  understood.  
Among them,  particularly remarkable  are the so-called 
precursor phenomena taking place in a temperature window 
at  first-order  structural phase transitions 
\cite{Na,Ohshima,Shapiro,Seto,Kinoshita,Uezu,Waseda}. 
They are  caused by  equilibrium or metastable 
coexistence of  disordered  and 
ordered  phases on mesoscopic scales  
and the volume fraction of the ordered regions increaes 
with lowering of the temperature,       
 as unambiguously 
observed with  various experimental methods. 
Theoretically, phase ordering 
phenomena in solids have been studied  
 using  time-dependent Ginzburg-Landau 
  or  phase field models,  in which 
an order parameter is coupled to the elastic field in 
a  coarse-grained free energy functional \cite{Kha,Onukibook}.  
Simulations of  the model dynamics  
 have  been  powerful in understanding the formation 
of mesoscopic domain  structures.

Kartha {\it et al.} \cite{Kartha} have ascribed 
the  origin of  the observed tweed patterns \cite{Ohshima} 
to  quenched disorder imposed by
the compositional randomness. 
In their model,   the elastic 
modulus  $C'=(C_{11}-C_{12})/2$  for the tetragonal 
deformations consists of a mean value proportional to $T-T_0$ 
and a space-dependent 
random noise, where $T_0$ is the 
nominal transition temperature.   
This randomness then 
produces  quasi-static  
flluctuations of  the strains, but it remains unclear 
how such a  microscopic perturbation  
can lead to mesoscopic 
  domain structures and lattice distortions 
 above $T_0$.   On the other hand, 
Seto {\it et al}. \cite{Seto} 
proposed     an intrinsic (impurity-free) pinnng 
mechanism   stemming from   anharmonic elasticity. 
One of the present authors  demonstrated  that 
third  order   anharmonic 
elasticity   can   freeze  
 tetragonal domains in a disordered 
matrix in   two-dimensional (2D) 
 simulation  \cite{pre,Onukibook}. 
Moreover, some authors 
stressed relevance of the anisotropic 
lattice deformations in  producing 
multi-phase coexistence in  perovskite manganites 
(possibly together with the compositional 
randoness) \cite{Mills, Bishop}. 
For example,  Ahn {\it et al.} \cite{Bishop} have used 
 a  2D phenomenological model with   
a two-component order parameter 
 coupled to the tetragonal   strain, 
but their  model free energy contains only  terms  even  
with respect to the order parameter (improper 
coupling).  As a result, in their model, 
the  domains  in phase ordering 
grow  anisotropically 
 up to the system size without pinning.

In  phase transitions 
subject to elsticity, there are various 
intrinsic pinning mechanisms  \cite{Onukibook}. 
As such examples, 
we mention binary alloys 
\cite{Nishimori} and polymer gels \cite{Puri}, 
where  the elastic moduli depend on the composition
(elastic inhomogeneity)  serving to freeze  
  the domain growth.  
 In  gel-forming  polymeric systems, furthermore, 
randomness in the crosslinkage 
constitutes quenched  disorder. Although  
 the domain pinning itself can be induced by  the 
elastic inhomogeneity only, the crosslink disorder   
 produces  quasi-static composition 
fluctuations enhanced 
toward the volume-phase and sol-gel 
transitions  \cite{Onukibook} 
as detected by scattering experiments
\cite{Tanaka-h,Shibayama}. 
 We also mention 
hexagonal-to-orthorhombic transformations 
 \cite{Onukibook,hexa}, 
where the interfaces between  the ordered variants  
take  preferred orientations hexagonal 
with respect to the principal 
lattice axes. This geometrical constraint gives rise 
to pinning of the domains.

In this  paper, we will further 
 investigate the  impurity-free  
pinning mechanism  at structural phase 
transitions in the Ginzburg-Landau scheme. 
We will present  analytic results  supported by 
 extensive numerical calculations.
We will present  only 2D results for the mathematical simplicity.  
We suppose  solid films  
undergoing a square-to-rectangle transition  in the plane. 
Such phase transitions would be realized   
under uniaxial compression \cite{Onukibook,JT}.  
However,  in real epitaxial films, 
analyzing   elastic effects 
 is difficult,  
because  the displacement is 
fixed at the film-substrate boundary 
and the stress is free at the film-air 
interface \cite{Desai}.

The organization of this paper  is as follows. 
In Section 2, we will  set up  the Ginzburg-Landau  free energy 
and the dynamic equation in 2D. 
The elastic field will be expressed in terms of $\psi$ 
under the mechanical equilibrium condition. 
In Section 3, we will  present 
analytic theory on  two kinds of intermediate states at fixed 
volume, in which ordered domains appear  in a disordered matrix and 
take shapes of long  stripes  or 
lozenges.   In Section 4,  we will give 
simulation results in 2D 
under the fixed-volume condition, 
which will be compared with our theory. 
In Appendix B, we will derive 
the effective free energy in the strain-only theory.

\section{Theoretical Background }

For  ferroelastic transitions, 
one  theoretical 
approach has been 
to set up a nonlinear elastic free energy 
containing the  strains    
up to sixth-order terms (the strain-only theory) 
\cite{Kartha,pre,Barsch,Jacobs,Curnoe,Lookman},  where some 
strain components  constitute a multi-component  order parameter 
and the corresponding elastic 
modulus decreases  toward the transition. 
Another approach has been to intoduce 
a {\it true}  order parameter   $\psi$ 
different from the strains. 
If   $\psi$  is coupled to  the tetragonal 
or shear strains {\it properly} or in the bilinear form   
in the free energy, 
elastic softening follows  above the transition 
 and anisotropic domains appear in phase ordering. 
  For improper structural 
phase transitions  \cite{Onukibook}, on the other hand, 
 the square of  the 
order parameter  $\psi^2$ is 
coupled to the strains without elastic softening. 
Thus  the second approach has been used  
for  improper structural phase transitions \cite{hexa,WangK}. 
In this paper, we will  adopt the second approach 
in 2D in the presence of both proper and improper 
elastic couplings, while one of the present authors 
presented a strain-only theory to describe 
intermediate states in 2D \cite{pre}.

\subsection{Ginzburg-Landau model in 2D}

We assume that the free energy function 
$F=F\{\psi,{\bi u}\}$ depends on a one-component  
order parameter $\psi$ and the  displacement $\bi u$.  
We define  $\bi u$ as the elastic displacement 
from  the atomic position  
in a reference one-phase state in the disordered matrix.  
No dislocation will be assumed in 
the  coherent condition \cite{Onukibook}, where 
$\bi u$ is treated as a continuous variable.

The free energy $F$   consists of four  parts, 
\be 
{F} = \int  d{\bi r}\bigg[   {f}_0  +\frac{C}{2}|\nabla\psi|^2
  + {f}_{\rm el} + {f}_{\rm int}\bigg].  
\en 
where the space integral is within the solid. 
The   free energy density  
$f_0= f_0(\psi)$  is the chemical part  
of  $\psi$ given in the Landau expansion form,   
\be
f_0 =
\frac{\tau}{2}\psi^2 + \frac{{\bar u}}{4}\psi^4 +\frac{v}{6}\psi^6  , 
\en 
where $\tau$ depends on the temperature $T$ as 
\be 
\tau =A_0(T-T_0).
\en 
We retain the terms up to the sixth order  in $f_0$. 
The coefficients $A_0$, $v$, and $C$ are positive 
constants, and 
 $T_0$ is  the critical temperature 
in the absence of the elastic coupling (for $\bar{u}>0$). 
We call $\tau$ the reduced 
temperature. The other coefficients are assumed to be independent 
of $T$. 
For the mathematical simplicity, we assume the 
isotropic elasticity with homogeneous elastic moduli. 
Then the elastic energy density  
is of the harmonic form, 
\be  
{f}_{\rm el} =
\frac{K}{2}e_1^2 + \frac{\mu}{2}(e_2^2 + e_3^2), 
\en 
where $K$ is  the bulk modulus 
and $\mu$ is the shear modulus. 
The    $e_1$, $e_2$, and $e_3$ 
are the dilational, tetragonal, and shear strains, 
respectively.  In terms of the displacement 
vector $\bi u$ ,  they are defined by 
\begin{eqnarray}
e_1 &=& \nabla _x u_x + \nabla _y u_y,\nonumber\\
e_2 &=& \nabla _x u_x - \nabla _y u_y,\nonumber\\
e_3 &=& \nabla _y u_x + \nabla _x u_y,     
\end{eqnarray}
where $\nabla_x=\p/\p x$ and $\nabla_y=\p/\p y$. 
The third term $f_{\rm int}$ consists of 
 the following two interaction terms, 
\be 
f_{\rm int} =  -g e_2\psi    -\alpha  e_1 \psi^2. 
\en
Thus $\psi$ is 
{\it properly} 
 coupled to the tetragonal strain $e_2$ 
and {\it improperly} 
 coupled to the  dilational  
strain $e_1$. 
Without loss of mathematical generality we may assume  
\be 
g >0, \quad \alpha >0.
\en
If both are negative,   we  change 
${\bi u}$ to $ -{\bi u}$. 
For $g<0$  and $\alpha>0$ we  change 
${\psi}$ to $ -{\psi}$, while 
for  $g>0$  and $\alpha<0$ 
 the signs of $\bi u$ and $\psi$ are both reversed.

In the absence of the elastic  coupling ($\alpha=g=0$)  
\cite{Onukibook,Griffiths}, it is well-known that  
the transition is second-order for $\bar u>0$ 
and is first-order for $\bar u<0$ with 
 the  tricritical  point at  $\tau={\bar u}=0. $
In our problem,   the phase transition behavior  
is much altered by $f_{\rm int}$.   
The  coupling to $e_2$  ($g\neq 0$)  gives rise to  
anisotropic domains (even in the isotropic elasticity) and  
 that  to $e_1$   ($\alpha\neq 0$)  makes  the effective reduced 
temperature($=\tau-2\alpha e_1$)  
 inhomogeneous in the intermediate states 
\cite{Onukibook}.

We introduce the elastic stress tensor by 
\begin{eqnarray}
 \sigma_{xx} &=& Ke_1 + \mu  e_2 - \alpha  \psi^2- g \psi,\\ 
 \sigma_{yy} &=& Ke_1 - \mu  e_2 - \alpha  \psi^2+ g \psi, \\
\sigma_{xy} &=& \sigma_{yx} = \mu  e_3.  
\end{eqnarray} 
If we change $\bi u$ by a small amount $\delta{\bi u}$, 
the free energy $F$ changes  by 
$\delta F= \int d{\bi r} \sum_{ij}\sigma_{ij} \p 
\delta u_j/\delta x_j$. 
We  then have $\delta F/\delta u_i= -\sum_{j} \nabla_j \sigma_{ij}$, where $\psi$ is fixed in the functional derivative with respect to $\bi u$. 
In  this paper, $\bi u$ is determined 
from the mechanical equilibrium condition, 
\be 
\sum_{j} \nabla_j \sigma_{ij}=-\frac{\delta F}{\delta{u_i}}=0.
\en 
Then $\bi u$ becomes 
a functional of $\psi$ under given boundary conditions 
\cite{Kha,Onukibook}.

In dynamics,  the elastic field is assumed to 
 instantaneously  satisfy the mechanical condition (2.11), 
while  the order parameter  $\psi$ 
   obeys the relaxation equation,  
\be
 \frac{\partial \psi}{\partial t} = -\lambda_0  
\frac{\delta F}{\delta\psi} 
= -\lambda_0  [\hat{\mu}-C\nabla^2\psi].
\en  
The kinetic coefficient $\lambda_0 $ is  
assumed to be a constant.   
From Eqs.(2,2) and (2.6)  
$\hat\mu $ in Eq.(2.12) is written as 
\be 
\hat{\mu} = 
\tau \psi  +\bar{u}\psi^3+ v\psi^5   
-g e_2 -2\alpha e_1\psi. 
\en
If we integrate Eq.(2.12), 
 equilibrium or  metastable states 
can be reached at long times, where 
it holds the extremum condition, 
\be 
\frac{\delta F}{\delta\psi}= 
\hat{\mu}-C\nabla^2\psi=0. 
\en 
The gradient term is important in  the interface regions, 
while  we may neglect it and set 
$\hat{\mu}=0$ outside them.

\subsection{Elimination of the elastic field}

 We may apply 
an  average homogeneous strain    as   
$\av{\nabla_j u_i}=A_{ij}$,
where $A_{ij}$  is a  homogeneous tensor 
representing  affine deformation.  
Hereafter  $\av{\cdots}= \int d{\bi r}(\cdots)/V$ 
represents the space average 
in the solid, $V$ being the solid volume. 
The displacement may  be divided 
into the average and the deviation as 
\be
{ u}_i= \sum_j A_{ij}x_j +  \delta{u_i} .
\en 
The simplest boundary condition 
is to  impose  the  periodicity of the deviation  
$\delta {\bi u}$. 
Note that the space average of 
$\sum_{ij} \sigma_{ij} \nabla_i\delta u_j$ vanishes 
in this  case. Since 
$2f_{\rm el}=\sum_{ij} \sigma_{ij} \nabla_i u_j+\alpha e_1 \psi^2
+g e_2 \psi$, the total  free  energy 
$F$ in Eq.(2.1) becomes  
\be
{F}= \int d{\bi r}\bigg [ 
f_0   + \frac{C}{2}|\nabla\psi|^2 
 + \sum_{ij} \frac{A_{ij}}{2}  
\sigma_{ij} \bigg ]+F_{\rm e}  .   
\en 
Under the mechanical equilibrium condition (2.11),  
the  elastic part $F_{\rm e}$ is written as 
\be 
F_{\rm e}=\int d{\bi r}\bigg [-\frac{\alpha}{2}\psi^2 e_1 
-\frac{g}{2}\psi e_2 \bigg ]
\en

In this paper, we impose the periodic boundary condition on  
$\delta {\bi u}$  and $\psi$  in 
 the region $0<x<V^{1/2}$ and 
$0<y<V^{1/2}$  in 2D to investigate  the  
domain morphology  not affected by the boundaries. 
It is then convenient to   use  the Fourier 
transformation,  
$
\delta u_j({\bi r})= 
\sum_{\bi k} u_{j{\bi k}} e^{i{\bi k}\cdot{\bi r}},  
$  
where $j=x$ and $y$  in 2D. 
The mechanical equilibrium condition (2.11) may be rewritten 
in terms of $u_{j{\bi k}}$  as  
\bea 
&&Kik_xe_{1{\bi k}}
-\mu k^2 u_{x{\bi k}}= ik_x[\alpha \varphi_{\bi k} 
+g \psi_{\bi k}] ,\nonumber\\
&&Kik_ye_{1{\bi k}} 
-\mu k^2 u_{y{\bi k}}= ik_y[\alpha \varphi_{\bi k} 
-g \psi_{\bi k}], 
\ena 
where $e_{1{\bi k}}$ and $\psi_{\bi k}$ 
are  the Fourier components  of  $e_1$ and  $\psi$, respectively, 
and $\varphi_{\bi k}$ is that  of 
\be 
\varphi({\bi r})= \psi
^2({\bi r})- \av{\psi^2} , 
\en 
Here  we make  $\av{\varphi}=0$. 
The Fourier components of the strains $e_1$, $e_2$, and 
$e_3$ are calculated 
for $\bi{k}\neq \bi{0}$  as 
\bea 
&&\hspace{-1.5cm}Le_{1{\bi k}} = {\alpha}\varphi_{{\bi k}}+ 
{g}\psi_{{\bi k}}\cos 2\theta,\\
&&\hspace{-1.5cm}
Le_{2{\bi k}} = {\alpha}\varphi_{{\bi k}} \cos 2\theta + 
{g}\psi_{{\bi k}} 
\bigg[1+\frac{K}{\mu} \sin^2 2\theta\bigg] 
,\\
&&\hspace{-1.5cm} 
Le_{3{\bi k}} = {\sin 2\theta}\bigg[ 
{\alpha} \varphi_{{\bi k}}- \frac{
K}{\mu} g\psi_{{\bi k}} \cos 2\theta \bigg], 
\ena  
where  $L$ is the longitudinal modulus defined by  
\be 
L=K+\mu.
\en 
We set $\cos\theta=k_x/k$  and   $\sin\theta=k_y/k$, so  
\be 
\cos 2\theta= (k_x^2-k_y^2) /k^2, \quad 
\sin 2\theta=2k_xk_y/k^2, 
\en 
in Eqs.(2.20)-(2.22). Here it is convenient to introduce 
a variable $\chi({\bi r})$ defined by 
\be 
\nabla^2 \chi= (\nabla_x^2-\nabla_y^2)\psi 
\en 
with $\av{\chi}=0$. The Fourier component of $\chi$ 
satisfies  $\chi_{\bi k}= \psi_{\bi k} \cos 2\theta$ 
for ${\bi k}\neq {\bi 0}$.  From Eq.(2.20)  $e_1$ is expressed as 
\be 
e_1 = \av{e_1}+ (\alpha \varphi+ g\chi)/L.
\en 
In terms of $\chi$ and $\varphi$, 
 $F_{\rm e}$ in Eq.(2.17) is expressed as 
\bea 
F_{\rm e}&=& \int d{\bi r}\bigg [ -\frac{g}{2}\av{e_2} \psi 
-\frac{\alpha}{2}\av{e_1}\psi^2  
- \frac{g^2}{2\mu}\delta\psi^2 \nonumber\\
&&+\frac{g^2K}{2\mu L}\chi^2 
 -\frac{\alpha g}{L}\chi\varphi 
- \frac{\alpha^2}{2L}\varphi^2\bigg],  
\ena 
where  $\delta\psi=\psi-\av{\psi}$. 
Thus $ F_{\rm e}$  
consists of  contributions of the  first, 
second, third, and fourth 
orders with respect to  $\psi$, where 
 the coefficients depend  on the space averages 
$\av{\psi}$  and $\av{\psi^2}$.

\subsection{Dimensionless representation}

In this paper, we will show snapshots of 
 $\psi$ and the strains obtained in our simulations. 
They will be normalized  as 
$\psi/\psi_0$, $e_1/e_0$, and $e_2/e_0$, where 
$\psi_0$ and $e_0$ are  typical amplitudes given by  
\be
\psi_0 = \frac{g}{\sqrt{|\bar u| K}} ,\quad  
e_0 = \frac{g}{K}\psi_0  = \frac{g^2}{\sqrt{{|\bar u| K^3}}}, 
\en
in terms of the coefficients 
 $g$, $\bar u$, and $K$ in the free energy  $F$. 
We will set $\mu=K$ and  include  the case of ${\bar u}<0$. 
In terms of the coefficient $C$ of the gradient free energy, 
the typical spatial scale $\ell$ is  given by 
\be 
\ell= \sqrt{CK}/g.
\en 
These  expressions give  the typical free energy density $C\psi_0^2/\ell^2= 
|{\bar u}|\psi_0^4= ge_0\psi_0=Ke_0^2= 
g^4/|\bar{u}|K^2$. We 
scale  the coefficients $\tau$, $\alpha$,  and $v$  by  
$\tau_0$,   $\alpha_0$, and $v_0$, respectively,    defined by 
\be
 \tau_0  = \frac{g^2}{K}  ,\quad  
\alpha_0 = {\sqrt{|\bar u| K}}, \quad  
v_0 = \frac{\bar u^2K}{g^2}. 
\en
Our simulation 
results  will  be  parametrized by  
$\tau/\tau_0$,  $\alpha/\alpha_0$, and $v/v_0$. 
If $\alpha\sim\alpha_0$ and $\psi\sim\psi_0$, 
the ratio of 
the typical dilational  strain ($\sim \alpha \psi^2/K$) 
to the typical tetragonal strain  ($\sim g \psi/\mu$) 
is of order $\mu/K(=1$ in our simulation). 
Near the tricritical point,  
$\bar u$  tends to  zero and  
 $\alpha_0$ and $v_0$  
become  small so that the scaled 
parameters  $\alpha/\alpha_0$ 
and $v/v_0$ are amplified.

\begin{figure}
\begin{center}
\includegraphics[width=8cm]{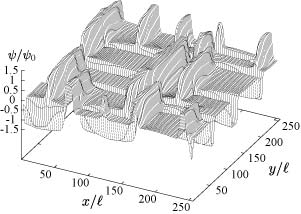}
\end{center}
\caption{Normalized order parameter 
$\psi/\psi_0$  in a  stripe  intermediate state 
 for $(\tau/\tau_0, \alpha/\alpha_0, v/v_0) 
=(0.6, 1.4,1)$. In the domains $\psi =\psi_{\rm s}$, $-\psi_{\rm s}$, 
or $0$.}
\end{figure}
\begin{figure}
\begin{center}
\includegraphics[width=6cm]{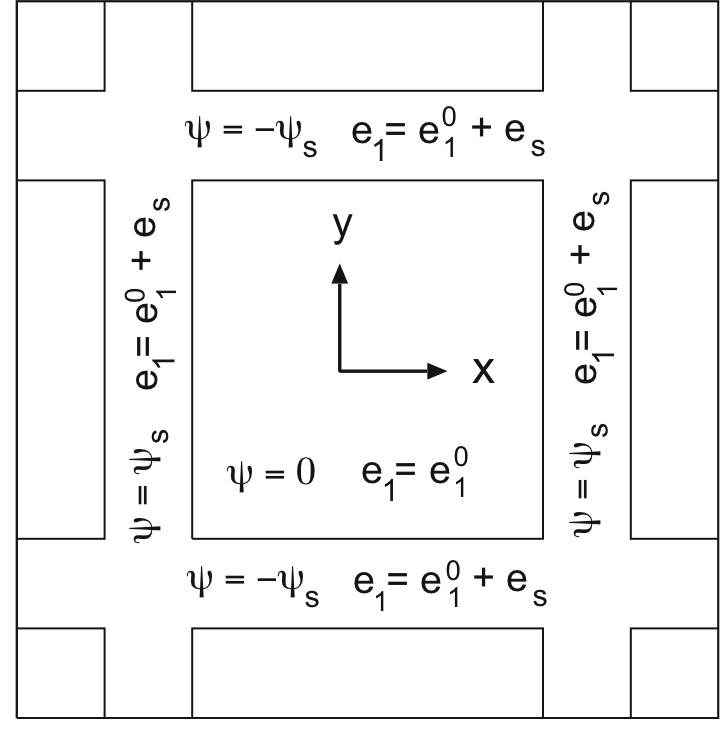}
\end{center}
 \caption{Disordered square regions 
are enclosed by ordered stripe domains 
uniaxially stetched along the $y$ or $x$ axis.}
\end{figure}

\begin{figure}
\begin{center}
\includegraphics[width=8cm]{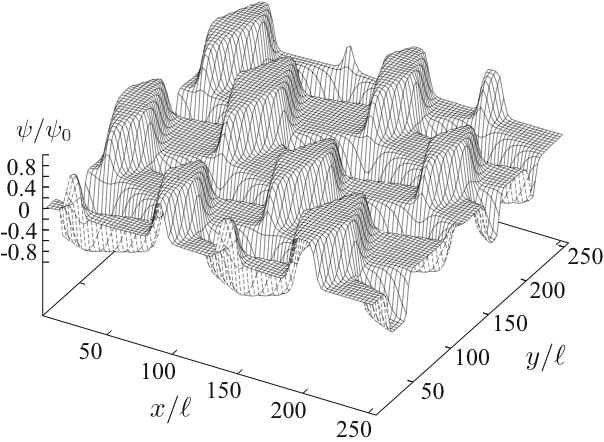}
\end{center}
\caption{Normalized order parameter 
$\psi/\psi_0$  in a  lozenge  intermediate state 
 for $(\tau/\tau_0, \alpha/\alpha_0, v/v_0) 
=(0.6,1,1)$. In the domains $\psi =\psi_\ell$, $-\psi_\ell$, 
or $0$.}
\end{figure}

\begin{figure*}
\begin{center}
\includegraphics[width=16cm]{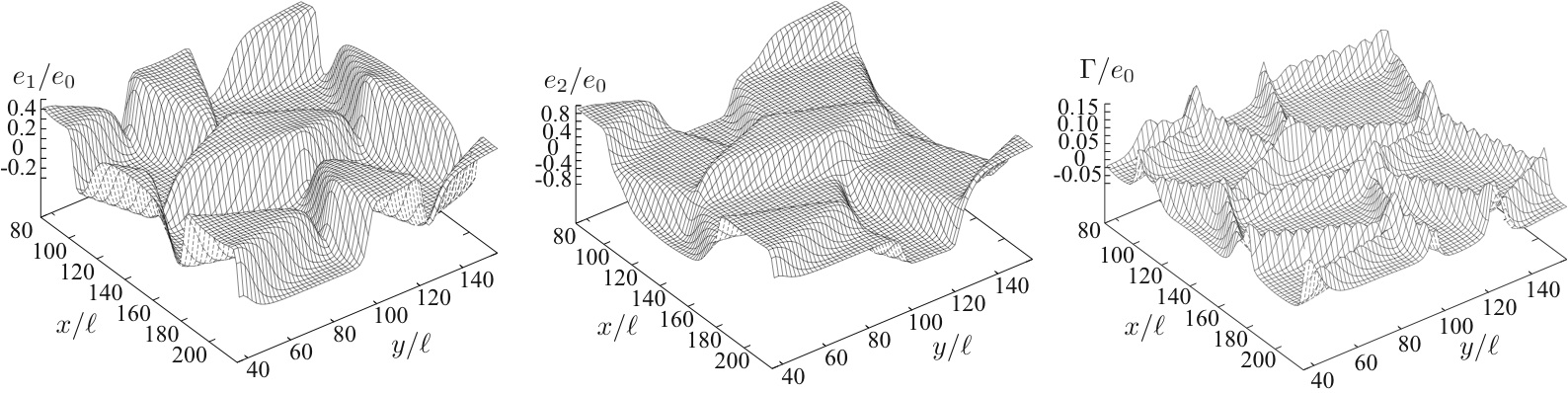}
\end{center}
\caption{Strains $e_1$, $e_2$, and $\Gamma=e_1-\alpha\varphi/K$ 
in units of $e_0$ in Eq.(2.28)  in the same lozenge 
state as  in Fig. 3. 
These stains are flat   and   
 $\Gamma$ is small within the domains. 
As a result, $e_1=  \alpha \varphi/K$, 
$e_2=g \psi/\mu$ in accord with Eq.(3.35).  
}
\end{figure*}

\section{Phase transitions at fixed volume and shape} 
\setcounter{equation}{0}

The phase transitions in solids  can 
crucially depend on the boundary condition \cite{Onukibook,comp}. 
In this paper, we limit ourselves to  the simplest case of 
 fixed volume (area in 2D) and   shape. 
This condition holds in the lateral directions 
if a solid film is fixed to a substrate.  
For 3D solids, the phase transition is 
usually observed  under the 
stress-free boundary condition 
and  claming of the boundaries 
is needed to realize the fixed-volume condition.

In this paper,  we thus assume  $A_{ij}=0$ ($i,j=x,y$) and 
\be 
\av{e_1}=\av{e_2}=\av{e_3}=0,
\en  
Furthermore, in our 2D simulation, there was    
no macroscopic order or 
\be 
\av{\psi}=0.
\en 
Then $\av{\sigma_{xx}}= \av{\sigma_{yy}}=-\alpha\av{\psi^2}$ 
and  $\av{\sigma_{xy}}=0$. Note that $\av{\psi}$ becomes 
nonvanishing under applied 
tetragonal strain $\av{e_2}\neq 0$, since  $\av{e_2}$ 
plays a role of the ordering field in Eq.(2.27). 
In  phase ordering,  
the condition (3.2) means that domains 
with positive $\psi$ and those with negative $\psi$ 
equally appear  and the space average of $\psi$ 
over many domains vanishes. Under Eqs.(3.1) and (3.2)  
the elastic contribution 
$F_{\rm e}$ in Eq.(2.17)  becomes 
\be 
F_{\rm e}= \int d{\bi r}\bigg [   
- \frac{g^2}{2\mu}\psi^2 +\frac{g^2K}{2\mu L}\chi^2 
 -\frac{\alpha g}{L}\chi\varphi 
- \frac{\alpha^2}{2L}\varphi^2\bigg],  
\en 
where $\varphi$ and $\chi$ are defined by 
Eqs. (2.19) and  (2.25), respectively.

In our previous paper \cite{comp}, 
the model  with $\alpha>0$ and $g=0$ has been 
studied  at fixed volume. 
There,   in a temperature window,   
the free energy is lower 
in  two-phase states than in  one-phase states, 
where the domains attain  macroscopic sizes, however.   
For   $\alpha=0$ and $g>0$,   on the other hand, 
lamellar or twin ordered states appear 
at low temperatures, where 
the interface normals make 
angles of $\pm \pi/4$ wth respect to 
the $x$ axis  and $\chi$ 
vanishes in 2D \cite{Barsch,pre,Jacobs,Curnoe,Lookman,Seme}.

\subsection{Stripe intermediate  states}

Figure  1 is an example of  
intermediate  states 
consisting of ordered stripes 
parallel to the $x$ or $y$ axis. 
As   illustrated in Fig. 2, 
 the stripes enclose 
disordered rectangular regions with  $\psi=0$  
isotropically dilated at 
$e_1= e_1^0$. 
A rotation of $\pi/2$  of a stripe pattern 
yields another possible stripe pattern. 
As a result, the areal 
fractions of the stripes along the 
$x$ axis and those along 
the $y$ axis are the same under no applied tetragonal strain.  
We assume that  the width of the stripes 
is much longer than the interface thickness.  
Within the stripes along the $y$ axis,  
we have  
\be 
\psi=\psi_{\rm s}, \quad 
e_1=e_1^0+ e_{\rm s},\quad   e_2= e_{\rm s}. 
\en 
On the other hand, within the stripes along the $x$ axis,   
$\psi$  and $e_2$ are reversed 
in sign, but $e_1$ is unchanged,    so that  
\be 
\psi=- \psi_{\rm s}, \quad 
e_1= e_1^0+e_{\rm s}, \quad e_2= -e_{\rm s} .  
\en  
 Thus we have $\av{\psi}=0$ and $\av{e_2}=
\av{e_3}=0$ on the average. From the dilation-free 
 condition $\av{e_1}=e_1^0+\phi e_s=0$ in Eq.(3.1),  
the areal fraction of 
the ordered regions $\phi$ is expressed as 
\be 
\phi=-e_1^0/e_s.
\en

To calculate the excess dilation 
 $e_{\rm s}$, we consider the interface region 
of a stripe along the $y$ axis, where 
all the quantities depend only on $x$. 
The mechanical equilibrium condition  yields 
$\sigma_{xx}=$constant along the $x$ axis. 
Here $\nabla_x u_x $ changes along the $x$ axis, while 
$\nabla_y u_y$ remains at $e_1^0/2$. 
From Eq.(2.8) or from Eq.(2.26) 
$\nabla_x u_x$ across 
the interface  is expressed as  
\be
\nabla_x u_x -\frac{1}{2}e_1^0= 
 \frac{\alpha}{L} \psi(x)^2+ \frac{g}{L}\psi(x), 
\en   
where $L$ is defined in Eq.(2.23). 
The above quantity is equal to 
$e_1(x)-e_1^0$ and $e_2(x)$. 
The result (3.7) also  follows  from 
Eqs.(2.20) and (2.21), 
where $\cos 2\theta= 1$ and $\chi=\psi$ since 
$\psi$ change along the $x$ axis.  The order parameter 
$\psi(x)$ changes from 0 to $\psi_s$ with increasing $x$, 
leading to  
\be  
e_{\rm s}= (\alpha\psi_{\rm s}^2+ g\psi_{\rm s})/L,  
\en    
In the  extremum condition (2.14) for $\psi(x)$, 
we set  $e_1-e_1^0$ and $e_2$ equal to the right hand side of 
Eq.(3.7) to obtain  
\be 
C\frac{d^2\psi}{dx^2}= 
f_0' -2\alpha e_1^0 \psi  
-(g  +2\alpha \psi)(\alpha\psi^2+ g\psi)/L, 
\en
where $f_0'=\p f_0/\p\psi$. This equation is integrated to 
give the standard form of the interface equation \cite{Onukibook}, 
\be 
\frac{C}{2} \bigg (\frac{d\psi}{dx}\bigg )^2= 
f_{\rm st}(\psi), 
\en  
where  
$f_{\rm st}(\psi)$ is 
 the effective free energy density near the interface. 
Here the derivative $\p f_{\rm st}(\psi)/\p \psi$ 
is equal to the right hand side of Eq.(3.9) 
and   $f_{\rm st}(0)=0$ is required,  
so we obtain     
\bea 
f_{\rm st} &=& f_0 -\alpha e_1^0 \psi^2 - (g+\alpha\psi)^2\psi^2/2L\nonumber\\
&=& \psi^2 \bigg [ 
\frac{\tau'}{2} + \frac{{ u}}{4}\psi^2 +\frac{v}{6}\psi^4 
-\frac{\alpha}{L}g\psi \bigg], 
\ena 
In the second line,   
 we define the coefficients,  
\bea 
\tau' &=&  \tau -g^2/L -2\alpha e_1^0, \\
\beta_L&=& 2\alpha^2/L,\\
u&=& {\bar u}- 2\alpha^2/L={\bar u}- \beta_L  .
\ena  
The integrand of $F_{\rm e}$ in Eq.(2.27) 
yields  the above $f_{\rm st}$ if we replace  $\av{e_1}$ 
by $e_1^0$, $\delta\psi$ by $\psi$, 
$\chi$ by $\psi$, and   $\varphi$ by $\psi^2$, 
 with $\av{e_2}=0$.

For the existence of a planar interface 
 $f_{\rm st}(\psi)$ should be minimized at the two points 
$\psi=0$ and $\psi_{\rm s}$, so we need to require 
$f_{\rm st}=\p f_{\rm st}/\p\psi=0$ at $\psi=\psi_{\rm s}$. Hence 
 $\psi_{\rm s}$ is the solution of the 
following cubic equation, 
\be 
\bigg(\frac{\p}{\p \psi}\frac{f_{\rm st}}{\psi^2}\bigg)_{\psi=\psi_{\rm s}} 
=\frac{{ u}}{2}\psi_{\rm s} +\frac{2v}{3}\psi_{\rm s}^3 
-\frac{\alpha}{L}g=0 . 
\en 
where  $\tau$ vanishes and $\psi_{\rm s}$ turns out to be 
independent of $\tau$ or 
the temperature.   
In Appendix A, we will explicitly solve the above 
cubic equation. From $f_{\rm st}(\psi_{\rm s})=0$ 
the effective reduced temperature 
 $\tau'$ is related to $\psi_{\rm s}$ by  
\bea 
\tau' &=& \frac{2\alpha}{L}g\psi_{\rm s} - \frac{{ u}}{2}\psi_{\rm s}^2 -\frac{v}{3}\psi_{\rm s}^4 ,\nonumber\\
&=&\frac{\alpha}{L}g\psi_{\rm s} + \frac{{ v}}{3}\psi_{\rm s}^4, 
\ena 
where the quadratic  term in the first 
line is eliminated in the second line with the aid of Eq.(3.15). 
Thus  $\tau'$ is also  a constant  
independent of $\tau$. 
With Eqs.(3.13) and (3.14), 
 we may rewrite $f_{\rm st}$ as 
\be 
f_{\rm st}=\psi^2(\psi-\psi_{\rm s})^2\bigg[\frac{v}{6}(\psi+\psi_{\rm s})^2+  
\frac{v}{3}\psi_{\rm s}^2+ \frac{u}{4}\bigg].
\en 
Since $f_{\rm st}(\psi)>0$ should hold at 
$\psi=-\psi_{\rm s}$, the inequality  
$v\psi_{\rm s}^2/3 + u/4>0$ follows, 
which then gives  $\psi_{\rm s}>0$ from Eq.(3.15). 
The positivity $\tau'>0$ also holds.

From Eq. (3.12) we have 
$e_1^0=(\tau -g^2/L-\tau')/2\alpha$. 
Use of Eqs. (3.8) and (3.16) gives the areal fraction 
$\phi$ in Eq.(3.6) in the form, 
\be 
 \phi  =  (\tau_{\rm s}-\tau)/\tau_{\rm sw}. 
\en 
Here 
$\tau_{\rm s}= \tau'+ g^2/L$ and $\tau_{\rm sw} = 
{2\alpha}e_{\rm s}$ are rewritten  as  
\bea  
&&\tau_{\rm s} = \frac{g^2}{L}  +
 \frac{\alpha}{L}g\psi_{\rm s} + \frac{{ v}}{3}\psi_{\rm s}^4,\\ 
&&\tau_{\rm sw} =  \frac{2}{L}\alpha\psi_{\rm s}(g+\alpha\psi_{\rm s}).
\ena 
Since  $\phi \rightarrow 0$ as $\tau\rightarrow \tau_{\rm s}$, 
 $\tau_{\rm s}$   is the  transition-point  value of $\tau$. 
From  $0<\phi <1$, 
$\tau$ is in the window range, 
\be 
\tau_{\rm s}-\tau_{\rm sw} <\tau < \tau_{\rm s},
\en 
to  realize the stripe domain morphology.   
For  $\tau\le \tau_{\rm s}-\tau_{\rm sw}$ the system  changes into an ordered 
state (where   
 the  order parameter 
and the free energy will be calculated  as   
Eqs.(3.47) and (3.48), respectively).

In the stripe intermediate states, 
the free energy density in the 
disordered regions is  $f_{\rm el}= K(e_1^0)^2/2$, 
while that in the stripes is  
$f_0+f_{\rm el}+f_{\rm int}= 
Ke_1^0( e_1^0/2+ e_{\rm s})$  
from $f_{\rm st}(\psi_{\rm s})=0$ or 
from Eq.(3.16), where $ e_{\rm s}$  is the excess dilation 
 given in  Eq.(3.8).  Neglecting  
the surface contribution, we write the  
 free energy density   as 
\bea 
\frac{F}{V}  &=& \frac{K}{2}e_1^0 (e_1^0+ 2\phi e_{\rm s})\nonumber\\ 
&=& -\frac{K}{8\alpha^2} (\tau_{\rm s}-\tau)^2, 
\ena 
where  the first line holds for general $e_1^0$ 
 and the second line 
follows from Eq.(3.6)  at fixed volume. 
Thus $F$ is negative in the stripe intermediate phase.

\subsection{Lozenge intermediate   states} 

\begin{figure}
\begin{center}
\includegraphics[width=8cm]{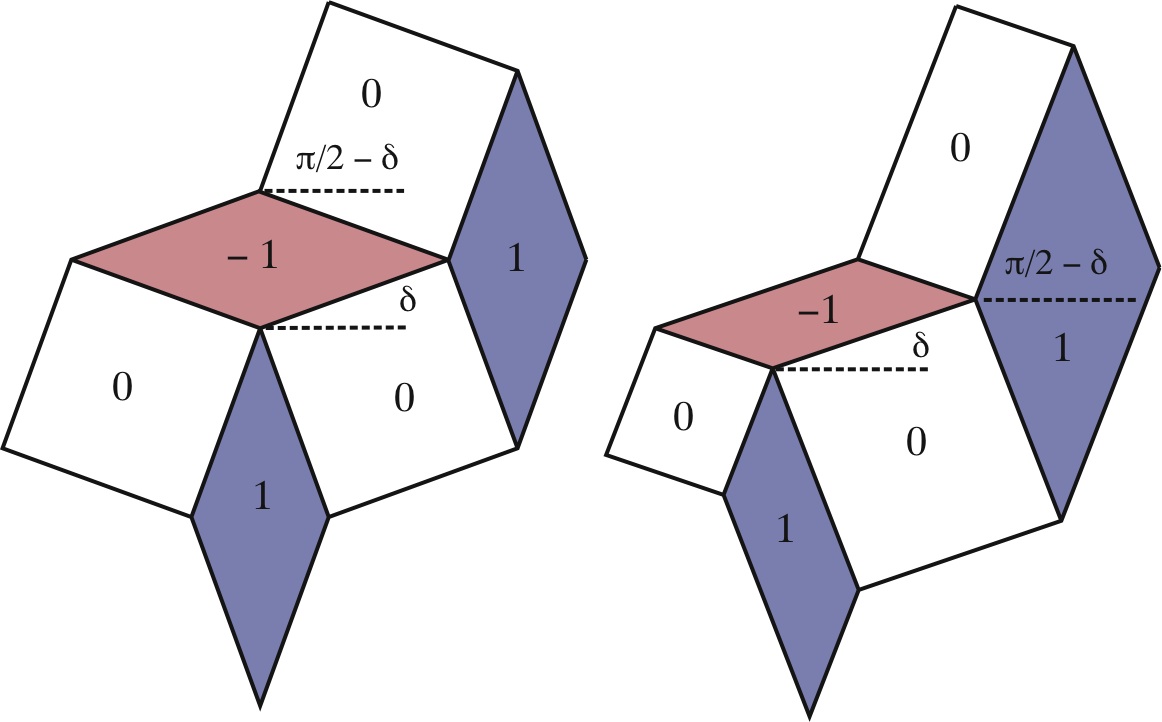}
\end{center}
\caption{Illustration of the lozenge geometry. 
Symmetric pattern consisting of 
lozenges and squares (left) 
and asymmetric pattern consisting of 
parallelograms  and rectangles (right).  
The angles  of the interface normals with respect to 
the horizontal axis are  written in terms of the angle $\delta$. 
The numbers $1$, $-1$, and 0 denote 
the value of $\psi(x,y)/\psi_\ell$. 
}
\end{figure}

\begin{figure}
\begin{center}
\includegraphics[width=8cm]{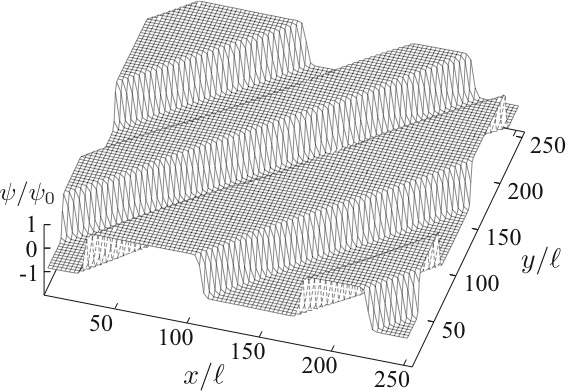}
\end{center}
\caption{Normalized order parameter 
$\psi/\psi_0$  in a twin   state 
consisting of two ordered variants separated by interfaces 
 for $(\tau/\tau_0, \alpha/\alpha_0,   v/v_0,) 
=(-1,1,1)$, where $\psi_0$, $\tau_0$, $\alpha_0$, and $v_0$ 
are given in Eqs.(2.28) and (2.30) and the spatial scale is $\ell $ in Eq.(2.29).  
}

\end{figure}

\begin{figure}
\begin{center}
\includegraphics[width=8cm]{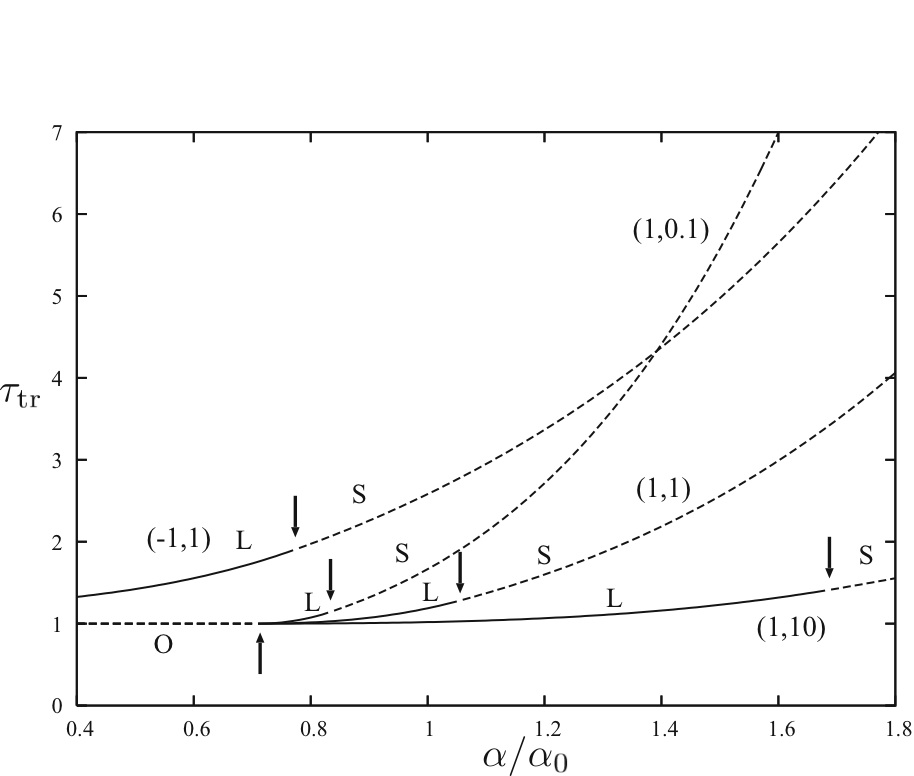}
\end{center}
 \caption{Reduced transition  temperature 
$\tau_{\rm tr}$ 
(maximum of    $\tau_{\rm s}$, 
 $\tau_\ell$, and $\tau_{0c}$)  
in units of $\tau_0$ 
as functions of $\alpha/\alpha_0$. 
The numbers to each curve represent  
$(\bar{u}/|\bar{u}|, v/v_0)$. 
The transition is from disordered to lozenge  
states on the solid lines (L). 
At the upward arrow  $\delta\rightarrow \pi/4$,  
and on its left (O)
the transition is to twin  ordered states. 
At the downward arrows  $\delta\rightarrow 0$,  
and on their right (S) the transition is to 
stripe states.}
\end{figure}

We show another kind of intermediate states. 
In Fig, 3, the pattern  consists of ordered  lozenges 
(rhombus) with $\psi=\pm \psi_\ell$ and disordered squares 
 with $\psi=0$.  As in Fig. 4, the strains 
 $e_1$ and $e_2$ change mostly 
in the interface regions 
and  are nearly flat (homogeneous) 
within each domain. 
The geometry is  illustrated in  Fig. 5. 
The ordered domains are characterized 
by the lozenge angle $\delta$ 
in the range $0<\delta<\pi/4$. More generally, 
as shown on the right,  the pattern may 
consist of ordered 
parallelograms (quadrilaterals with 
both pairs of opposite sides being  parallel)
and disordered rectangles. 
The stepwise behavior of the strains 
in each domain in Fig. 4 is  a characteristic feature 
of the lozenge geometry in Fig.5. 
We may  derive it starting with the assumption 
of  the stepwise behavior of $\psi$.  
We  first examine  the behavior 
of $\chi$ in Eq.(2.25)  
in  the vicinity of an interface 
between  ordered and  disordered domains. 
Except near the corners 
of the lozenges, we may set 
\be 
\psi(x,y) =\pm \psi_\ell \Psi(\zeta),
\en  
where $+$ (or $-$) is for the ordered 
phase with $\psi=\psi_\ell$  (or 
$\psi=-\psi_\ell$) and  
$\zeta$ is the coordinate along 
the interface 
normal direction ${\bi n}=
(n_x,n_y)$. The  function $\Psi(\zeta)$ 
tends to  unity  in the interior of 
all the ordered domains, while it vanishes 
in the disordered regions. From Eq.(2.25) we obtain 
$
{d^2\chi}/{d\zeta^2}= \pm \cos(2\theta_0)\psi_\ell  
 {d^2\Psi}/{d\zeta^2}$, where  $\cos(2\theta_0)=n_x^2-n_y^2$ 
with $\theta_0$  being   the angle 
between $\bi n$ and the $x$ axis. 
From Fig. 5 we can see that 
$\theta_0= \pm \delta$ (or $\theta_0= \pi/2\pm \delta$) 
for the ordered phase  with $\psi=\psi_\ell$ 
(or  for that with $\psi=-\psi_\ell$) so that 
 $\cos(2\theta_0)= \pm \cos(2\delta)$. 
Thus   we find 
\be 
\frac{ d^2\chi}{d\zeta^2}= \cos(2\delta) \psi_\ell  
 \frac{d^2\Psi}{d\zeta^2},
\en  
where the symbol $\pm$ has disappeared.  
This is satisfied if $\chi- \cos(2\delta)\psi_\ell\Psi$ 
is a constant or if 
\be 
\chi(x,y)  =  [\Psi(x,y) -\phi]\psi_\ell \cos(2\delta),
\en  
where $\phi=\av{\Psi}$  is 
 the areal  fraction of the ordered regions 
and $\av{\chi}=0$  is satisfied 
(since $\Psi(x,y)$ is equal to 1 in the ordered regions 
and to 0 in the disordered regions). 
We  numerically calculated $\chi(x,y)$ 
 from the inverse Fourier transformation of  $\chi_{\bi k}= 
(k_x^2-k_y^2)\psi_{\bi k}/k^2$ to  
confirm  the above relation for 
a number of the lozenge states,  
where the agreement becomes better 
with decreasing the interface width  
(or decreasing  the coefficient $C$ in Eq.(2.1)).  
Similarly,  the strain $e_2(x,y)$ is proportional to 
$\psi(x,y)$ and is  of the form, 
\be 
e_2(x,y)= [g- K\Gamma_c\cos(2\delta)]\psi(x,y)/\mu. 
\en
where we define 
\be 
\Gamma_c= [g \cos(2\delta) 
- \alpha \mu\psi_\ell/K]\psi_\ell/L.
\en
Here the inverse Fourier transformation 
of $\alpha \varphi_{\bi k}\cos(2\theta)$ in Eq.(2.,21) 
has been  approximated  by $\alpha \cos(2\delta) 
\psi_\ell\psi(x,y)$. 
In the thin interface limit, 
$\psi(x,y)/\psi_\ell (=0, \pm 1)$ and 
$\Psi(x,y)(=0,1)$ are  step functions 
and  Eqs.(3.25) and (3.26) 
hold  exactly as solutions of Eqs.(2.21) and (2.25).

We rewrite the  elastic free energy in Eq.(3.3)  as 
\be 
F_{\rm e} = \int d{\bi r}\bigg [-\frac{g^2}{2\mu}\psi^2 
-\frac{\beta_K}{4}\varphi^2 
+ \frac{KL}{2\mu }{ \Gamma^2}\bigg ].
\en  
where  we define the coefficient $\beta_K $ 
and introduce the quantity $\Gamma$ by 
\bea 
\beta_K&=&2\alpha^2/K,\\
\Gamma&=& \frac{g}{L}\chi - \frac{\alpha\mu }{KL}\varphi.
\ena   
Then  $e_1= \Gamma+ \alpha\varphi/K$ in terms of $\Gamma$. 
From Eq.(3.24) we find  
$\Gamma=(1-\phi)\Gamma_c$ in the ordered regions and  
$\Gamma= -\phi\Gamma_c$ in the disordered regions with  
 $\Gamma_c$ being defined by Eq.(3.27), so that $\av{\Gamma}=0$ and 
$
\av{ \Gamma^2}= \phi(1-\phi)
\Gamma_c^2$.
In addition $\av{\varphi^2}= \psi_\ell^4\phi(1-\phi)$.  
Neglecting the surface tension contribution, 
we may calculate the average 
free energy density  as 
\be 
\frac{F}{V}= f(\psi_\ell) \phi +\bigg[ - 
\frac{\beta_K}{4} \psi_\ell^4 
+\frac{KL}{2\mu } { \Gamma_c^2}\bigg]\phi(1-\phi).  
\en       
For simplicity,  we introduce 
\be 
f(\psi) = f_0(\psi) -{g^2}\psi^2/2\mu, 
\en 
which is of the same form as $f_0$ in Eq.(2.2) with 
$\tau$ being shifted to $\tau-g^2/\mu$.

The average free energy in Eq.(3.31)  
 depends on $\delta$, $\psi_\ell$, and 
 $\phi$.   In our simulation to follow, 
we shall see that both the symmetric and asymmetric patterns 
in Fig. 5  both appear and  the edges of the 
ordered regions are  flattened with decreasing 
$\tau$  (Figs. 11 and 14). This means that $\delta$ and $\phi$ 
can be varied  independently, though they are 
related as $\phi=\sin(2\delta)/(1+\sin(2\delta))$ 
for the symmetric pattern (left of Fig.5).   
Thus  we minimize $F$ 
with respect to these three 
quantities. First, from Eq.(3.27) 
we notice that ${\Gamma_c}$ can  vanish 
for 
\be 
\alpha \mu \psi_\ell/gK<1,
\en 
if the angle $\delta$ is chosen as 
\be 
\delta=\frac{1}{2} \cos^{-1}(\alpha\mu \psi_\ell/gK).  
\en 
If we set $\Gamma=0$ or $\chi=(\alpha\mu/gK)\varphi$,  
we obtain 
\be 
e_1= \alpha \varphi/K, \quad e_2= g\psi/\mu,
\en 
where the former follows from 
Eq.(2.26) and the latter from 
Eq.(2.21) or from Eq.(3.36). 
The snapshots in Fig. 4 are consistent with 
the above relations  except in the 
interface regions.

Under Eqs.(3.34) and (3.35)  we 
 furthermore minimize $F/V$  with respect to 
 $\psi_\ell$ and $\phi$. 
If $\Gamma_c=0$, 
 the coefficient of the quadratic 
term $(\propto \phi^2)$ is positive in Eq.(3.31) and 
$F/V$ can take 
 a minimum as a function of $\phi$ in the range $0<\phi<1$ 
for sufficiently small  $f(\psi_\ell)$. 
From Eq.(3.31) the minimum conditions read 
\bea 
&&\hspace{-1cm}
\frac{1}{\phi V}\frac{\p F}{\p \psi_\ell}=f'(\psi_\ell)- \beta_K\psi_\ell^3(1-\phi) =0,\\  
&&\hspace{-1cm}
\frac{1}{V}\frac{\p F}{\p \phi}=
f(\psi_\ell) - \frac{1}{4} \beta_K \psi_\ell^4 (1-2\phi)=0, 
\ena
where $f'=\p f/\p \psi$.  The first equation 
(3.36) is equivalent to   the equilibrium condition 
(2.14) outside the interface regions 
if use is made of  Eq.(3.35). 
In our previous work on the compressible 
Ising model\cite{comp}, similar 
 mminimization   of $F$ 
yielded two-phase coexistence in a temperature window 
(leading to Eqs.(3.49)-(3.52) in the next subsection).
In the present case, we may eliminate 
 $\phi$ from Eqs.(3.36) and (3.37)  to 
obtain $\psi f' -2f - \beta_K \psi^4/2=0$ 
at $\psi=\psi_\ell$. This  is solved to give  
\be 
\psi_\ell^2= {3}(\beta_K- \bar{u})/4v,
\en 
from which we need to require $\beta_K>\bar{u}$. 
Remarkably,  $\psi_\ell$ becomes  independent of $\tau$ 
as well as $\psi_{\rm s}$ determined by  Eq.(3.15). 
From Eq.(3.36) the areal fraction $\phi$ is expressed as 
\be 
\phi= 1-f'(\psi_\ell)/\beta_K\psi_\ell^3= 
(\tau_{\ell}-\tau)/\tau_{\ell {\rm w}}, 
\en 
in the same form as in Eq.(3.18). Here,   
\bea 
&&\tau_\ell=\frac{g^2}{\mu} +  \frac{v}{3}\psi_\ell^4
= \frac{g^2}{\mu} + 
\frac{3}{16v} (\beta_K- \bar{u})^2,\\
&&\tau_{\ell {\rm w}} = \beta_K \psi_\ell^2
= \frac{3}{4v}  \beta_K (\beta_K- \bar{u}).
\ena 
The $\tau_\ell$ is  the transition reduced 
temperature and $\tau_{\ell {\rm w}}$  is 
the width of the window. 
The lozenge intermediate phase is  
realized in the  window, 
\be 
\tau_\ell-\tau_{\ell {\rm w}}< \tau < \tau_\ell.
\en    
As in Eq.(3.22) the average free energy density is calculated as 
\be 
\frac{F}{V} =    -\frac{K}{8\alpha^2} (\tau_\ell-\tau)^2. 
\en

Now we return to 
Eq. (3.33) and 
the inequality $\beta_K>\bar{u}$ implied by Eq.(3.38). 
They impose  the range of $\beta_K= 2\alpha^2/K$ 
allowed  for the lozenge phase, 
\be 
{\bar u}< \beta_K< \frac{\bar{u}}{2} + 
\sqrt{\frac{\bar{u}^2}{4}+ \frac{8g^2 }{3\mu^2}Kv}.
\en 
Here Eq.(3.33) gives 
the upper bound, at which   $\delta$ tends to 
$0$ and the lozenge 
patterns continuously change into those of 
stripes.  Therefore,  the lozenge-stripe boundary is determined by 
$\psi_\ell = gK/\mu\alpha$, at which  
  $\psi_{\rm s}=\psi_\ell$, leading to 
 $\tau_{\rm s}=\tau_\ell$ from Eqs.(3.19) and (3.40). 
The continuity of 
$F/V$ also holds from Eqs.(3.22) and (3.43).

\subsection{One-dimensionally  ordered states}

With decreasing $\tau$,  
our system  eventually consists of regions of  two 
ordered variants  with $\psi=\pm \psi_{\rm e}$ 
at  fixed volume and shape in steady states. 
In Fig. 6, we show  such a 
 lamellar or twin ordered state,  
where  the two ordered variants are 
separated by antiphase boundaries or twin walls  
and there is no bulk disordered region 
\cite{Barsch,pre,Jacobs,Curnoe,Lookman}. 
Here  we  assume that 
 all the quantities  depend  only 
on $x'= (x-y)/\sqrt{2}$ as in Fig.6. Then we obtain   
 $\chi=0$, leading to 
$e_1 =\alpha\varphi/L$,  and $e_2=g\psi/\mu$.
The resultant free energy at fixed volume is expressed as 
\be 
F= \int d{\bi r} \bigg [f 
+ \frac{C}{2} |{\psi'}|^2
- \frac{\beta_L}{4}\varphi^2\bigg],
\en 
where $\psi'= d\psi/dx'$ and $\beta_L$ is defined by   
Eq.(3.13). The  above  form coincides with 
 the free energy of 
the  compressible Ising model at fixed volume (with no 
ordering field conjugate to $\psi$)  
 \cite{comp}, on the basis of which 
we discuss the two cases below.

If  $u={\bar u}- \beta_L$ in Eq.(3.14) 
is positive, a  twin ordered state 
is realized for 
\be 
\tau<
\tau_{0c}= g^2/\mu   , 
\en 
where we may set $\varphi=0$ 
and then $\psi_{\rm e}$ is the solution of 
$f'(\psi_{\rm e})=0$ with  $f$ being  defined by Eq.(3.32). 
It is solved to give   
\be 
\psi_{\rm e}^2= [{\bar u}^2 -4v(\tau -g^2/\mu)]^{1/2}/2v -{\bar u}/2v.
\en 
The elasticity effect is  only to 
shift  $\tau$ to $\tau-g^2/\mu$. 
The average free energy 
density  takes a negative value,    
\be
F/V =-\psi_{\rm e}^4( 3\bar{u}+4v\psi_{\rm e}^2)/12 ,
\en 
where the surface tension contribution is neglected.

\begin{figure*}
\begin{center}
\includegraphics[width=16.0cm]{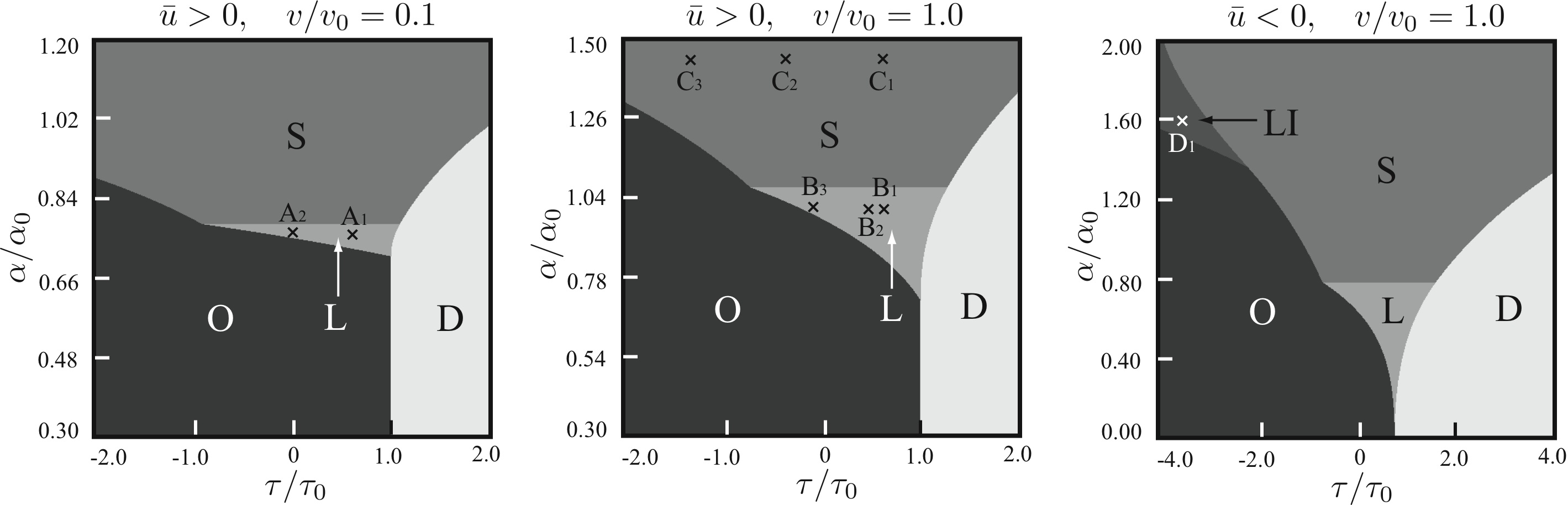}
\end{center}
\caption{Theoretical phase diagrams 
in the plane of $\tau/\tau_0$ and $\alpha/\alpha_0$  for 
$({\bar u}/|{\bar u}|, v/v_0)= (1,0.1)$ (left), 
$(1,1)$ (middle), and $(-1,1)$ (right). 
For ${\bar u}>0$, we  have the disordered  (D), 
twin  ordered  (O), stripe intermediate (S), 
and lozenge intermediate (L) phases. For ${\bar u}<0$ (right),  
there appears the lamellar intermediate phase (LI), where 
the disordered and ordered regions coexist macroscopically \cite{comp}. 
The points A$_1$, B$_1$, B$_2$, 
C$_1$, C$_2$, and C$_3$ represent  intermediate states 
with patterns shown in Figs. 9-13. At the points 
A$_2$ and B$_3$  twin  patterns 
were realized numerically.
}
\end{figure*}

For $u={\bar u}- \beta_L<0$,  on the other hand, 
the elastic term proportional to $\beta_L$ in Eq.(3.45)  
can give rise to macroscopic 
 coexistence of 
disordered  and 
 ordered regions 
in the  following window, 
\be 
\tau_{0c}' -  \tau_{0w}  < 
\tau<\tau_{0c}', 
\en 
where  $\psi= \pm (3|u|/4v)^{1/2}$ 
in the ordered  regions. 
The transition-point value and the window width 
in this case are given by 
\bea 
\tau_{0c}'&= & \frac{g^2}{\mu} +  
\frac{3{u}^2}{16v}=\frac{g^2}{\mu} + 
\frac{3}{16v} (\beta_L- \bar{u})^2  ,\\
\tau_{0w}&=& \frac{3\beta_L }{4v}|u| 
= \frac{3}{4v}  \beta_L (\beta_L- \bar{u}).
\ena   
The areal fraction of the ordered 
regions is $\phi= (\tau_{0c}' -\tau)/\tau_{0w}$.  
The average free energy  
without the surface tension  contribution becomes   
\be 
\frac{F}{V}= -\frac{L}{8\alpha^2} (r_{0c}' -r)^2 .
\en 
Below the lower bound of the window in Eq.(3.49) 
the twin ordered phase with  $\psi =\pm \psi_{\rm e}$ is realized. 
In our simulation,  however, we did  not realize 
the macroscopic coexistence predicted above.

\subsection{Theoretical phase behavior}

As  $\tau$ is lowered from a  value in a 
disordered state below a 
 reduced  transition temperature $\tau_{\rm tr}$, 
phase ordering  occurs into an intermediate  or 
ordered state. 
Depending on which is  largest,  $\tau_{\rm tr}$ 
is given by either of $\tau_{\rm s}$ in Eq.(3.19), 
$\tau_{\ell}$ in 
Eq.(3.40)  under  Eq.(3.44), $\tau_{0c}$ in 
Eq.(3.46) for $u>0$,  
or $\tau_{oc}'$ in Eq.(3.50) for $u<0$. 
In Fig. 7, we show it as a function of $\alpha/\alpha_0$ 
for four sets of $({\bar u}/|{\bar u}|, v/v_0)$ 
at $\mu=K$.   (i) For $\bar u>0$,  
the transition is at $\tau=g^2/\mu$ 
to the twin    phase for $\alpha/\alpha_0 < \sqrt{2}$, 
to the lozenge phase for the intermediate range,  
\be 
\sqrt{2} <\alpha/\alpha_0< \sqrt{1+ A},
\en  
where $A\equiv (1+ 32K^2v/3\mu^2v_0)^{1/2}$ from Eq.(3.44). 
For larger $\alpha/\alpha_0$  the stripe phase 
is realized. (ii) For ${\bar u}<0$, the transition is 
to the lozenge phase for $\alpha/\alpha_0
<\sqrt{1+ A}$ 
and to the stripe phase for $\alpha/\alpha_0
>\sqrt{1+ A}$.

\begin{figure}
\begin{center}
\includegraphics[width=8cm]{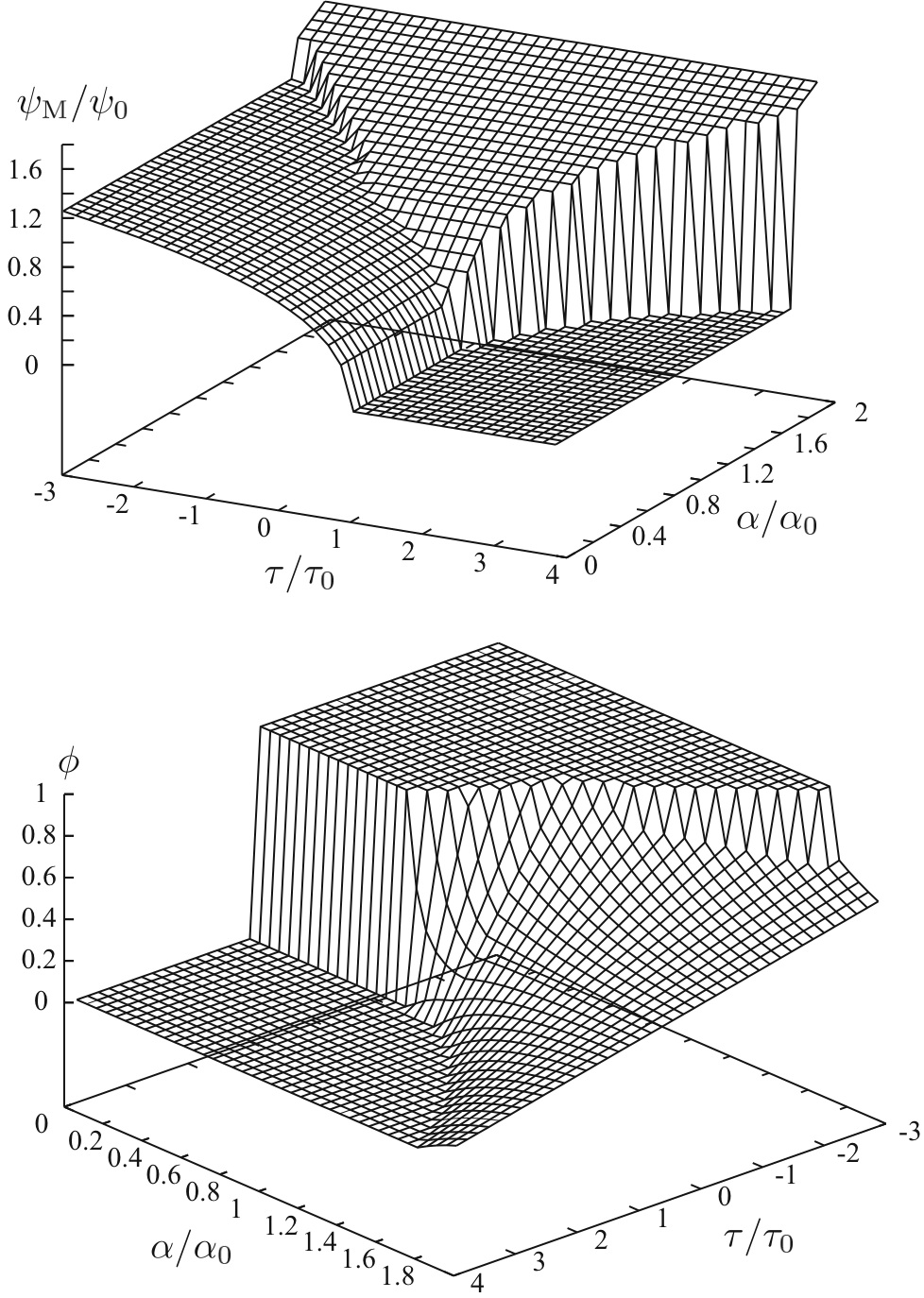}
\end{center}
\caption{Theoretical  order parameter 
$\psi_M$ scaled by $\psi_0$ (upper plate), 
which is $\psi_{\rm s}$, 
$\psi_\ell$,  or $\psi_{\rm e}$ in the stripe, 
lozenge, or twin  phase, respectively. 
Theoretical areal fraction $\phi$ of the ordered regions (lower panel) 
changing from 0 to 1 as $\tau$ is decreased.  
Here ${\bar u}>0$ and $v/v_0= 1$ and 
the corresponding phase diagram is 
 the middle plate of Fig. 8. 
}
\end{figure}

\begin{figure}
\begin{center}
\includegraphics[width=8cm]{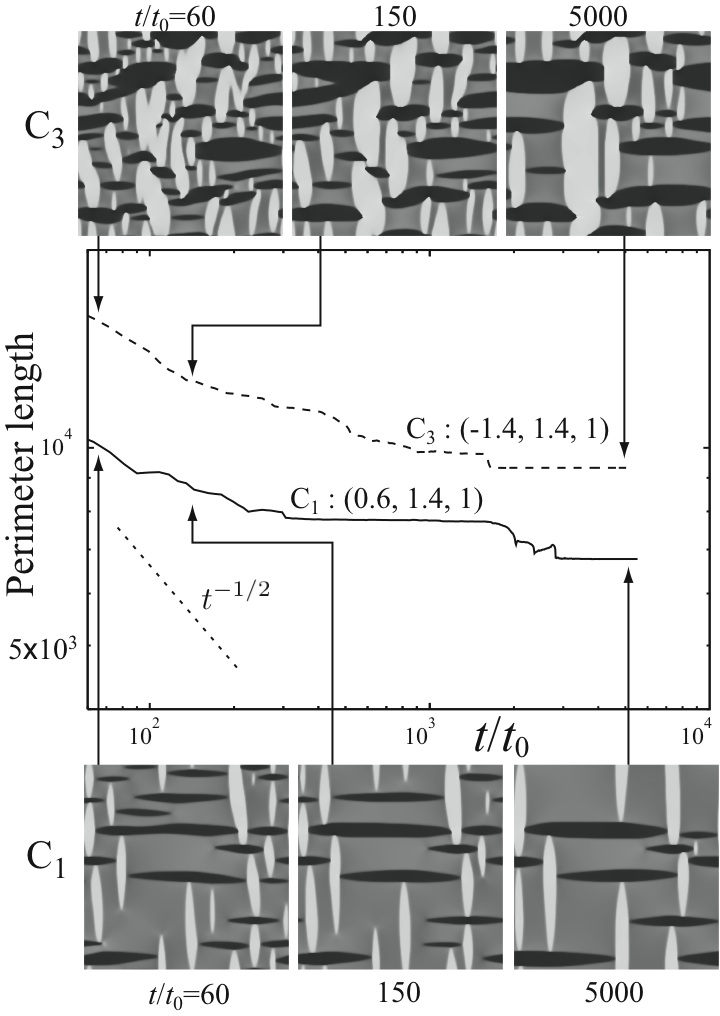}
\end{center}
 \caption{Relaxation of the total perimeter length 
 after quenching  from a disordered  to  stripe 
state at  C$_1$ (lower curve) and   C$_3$ (upper curve) 
in Fig. 8.  Evolution  of $\psi$ 
is displayed. Times are  in units of $t_0$ in 
Eq.(4.1).}
\end{figure}

\begin{figure}
\begin{center}
\includegraphics[width=8cm]{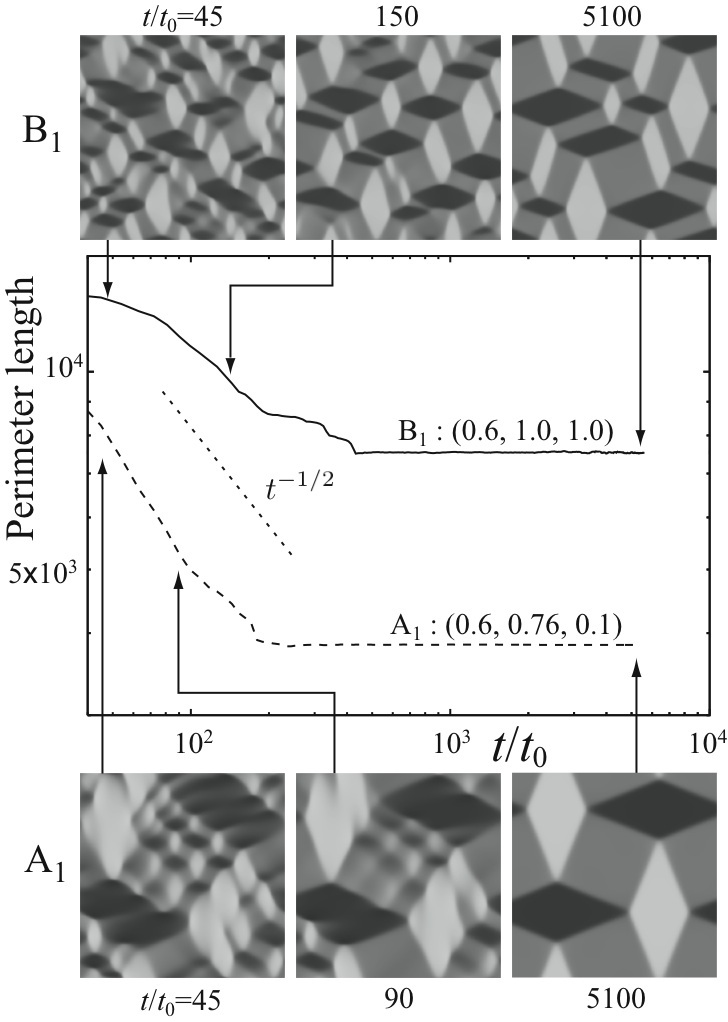}
\end{center}
 \caption{Relaxation   from a disordered   
 to  lozenge state  at A$_1$ (lower curve) and  B$_1$ 
(upper curve) in Fig. 8. The final  pattern 
is symmetric at  A$_1$ and asymmetric 
 at  $B_1$ (see Fig. 5). 
}
\end{figure}

\begin{figure*}
\begin{center}
\includegraphics[width=12cm]{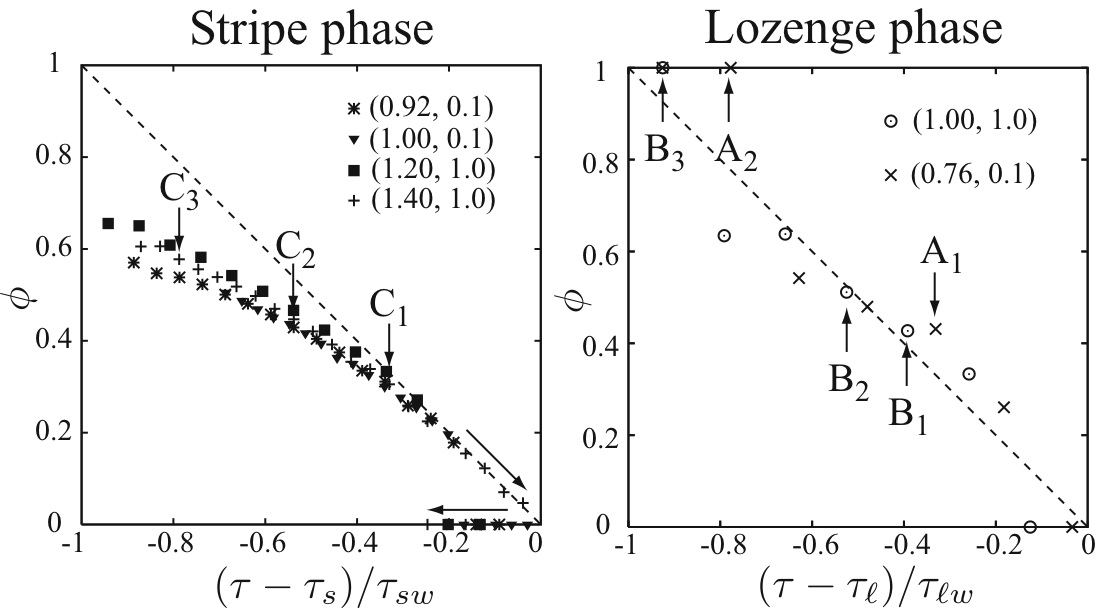}
\end{center}
\caption{Numerical areal fraction of the ordered regions  
$\phi$ as a function of $(\tau-\tau_{\rm s})/\tau_{\rm sw}$ 
(left) and  $(\tau-\tau_\ell)/\tau_{\ell {\rm w}}$ (right) 
in the   stripe and  lozenge  phases,  respectively. 
The numbers are the values of 
$(\alpha/\alpha_0,v/v_0)$.  See Fig. 8 for the positions of  
C$_1$, C$_2$,  C$_3$,  A$_1$, A$_2$, B$_1$, B$_2$, 
and B$_3$ in the phase 
diagrams. The arrows in the left panel indicate  how 
the data were obtained in the range $g^2/\mu<\tau<\tau_s$.}
\end{figure*}

We  may calculate a  phase diagram  in the plane of 
$\tau/\tau_0$ and $\alpha/\alpha_0$  using the scalings 
in Eqs.(2.28)-(2.30) if we  give  
$\mu/K$, $v/v_0$, and the sign of $\bar u$. 
 Figure 8 displays such examples at 
$({\bar u}/|{\bar u}|, v/v_0)= (1,0.1)$, 
$(1,1)$, and $(-1,1)$  at $\mu/K=1$. For ${\bar u}>0$, 
the  plane is divided into 
the disordered (D), twin  ordered (O), 
stripe (S), and lozenge (O) regions, depending on which morphology 
has the lowest free energy 
(the interface free energy being neglected). 
The lozenge region can appear 
for $\alpha/\alpha_0>\sqrt{2}$ 
from Eq.(3.44) and  is narrow for small $v/v_0$ 
(or far from the tricriticality).  
The boundary between the lozenge and twin  phases 
is determined by $\tau=\tau_\ell-\tau_{\ell w}$ 
(or $\phi=1$)). In fact, at this temperature, 
Eq.(3.47) gives $\psi_{\rm e}^2=3(\beta_K-\bar{u})/4v=\psi_\ell^2$ 
 and $F/V$ in Eq.(3.48) becomes 
$-K\tau_w^2/8\alpha^2$, 
demonstrating   the continuity of 
$\psi$ and $F/V$ at the boundary from Eqs.(3.38) and (3.52). 
For ${\bar u}<0$,  the lamellar  region 
is further divided into the ordered region (O)  and 
the intermediate region of 
macroscopic two-phase coexistence (LI) 
 described by Eqs.(3.49)-(3.52). The latter 
phase (LI) was not realized in our simulation, however.

Furthermore, in Fig. 9, we plot the equilibrium order parameter 
$\psi_M$ in the upper plate 
and the equilibrium areal fraction 
$\phi$ in the lower plate for 
${\bar u}>0$ and $v/v_0=1$.  The former  is given by  $\psi_{\rm s}$,  
$\psi_\ell$,  or $\psi_{\rm e}$ in the stripe, 
lozenge, or twin  phase, respectively. It 
is continuous at the disorder-twin  and lozenge-twin 
boundaries and discontinuous at the other phase boundaries.   
The $\phi$ is between 0 and 1 in the intermediate phases. It 
is continuous at the disorder-stripe, disorder-lozenge, 
and lozenge-twin  phase boundaries  and discontinuous 
at the other phase boundaries. 
We can  see that $\psi_M$ and $\phi$ 
are both discontinuous at the twin-stripe boundary.

\section{Numerical results}
\setcounter{equation}{0}

We  numerically integrated the dynamic equation (2.12) in 2D 
under the periodic boundary condition 
on a  $256\times 256$ lattice. 
We measure space and time 
in units of $\ell$ in Eq.(2.29) and 
\be 
t_0= \ell^2/\lambda_0C= K/\lambda_0 g^2,
\en   
where $\lambda_0$ is the kinetic coefficient. 
When we started with a disordered phase, 
the initial value of $\psi$ was a random number 
in the range $|\psi|<0.01\psi_0$ at each lattice point. 
No random source term was  added to the dynamic equation. 
We calculated the strains $e_1$ and $e_2$ 
 using their Fourier components in Eqs.(2.20) and (2.21)
 under the fixed volume condition.  We set 
$\av{e_1}=\av{e_2}=0$. 
We will show the resultant patterns 
at the points  marked by $\times$ 
 in the  phase diagrams in Fig. 8 (except  A$_2$  and B$_3$). 
The parameters $(\tau/\tau_0,\alpha/\alpha_0,v/v_0)$ 
are given by 
A$_1$$:(0.6,0.76,0.1)$, 
A$_2$$:(0,0.76,0.1)$, 
B$_1$$:(0.6,1,1)$,  
B$_2$$:(0.4,1,1)$, 
B$_3$$:(-0.2,1,1)$, 
C$_1$$:(0.6,1.4, 1)$,
C$_2$$:(-0.4,1.4,1)$,
 and C$_3$$:(-1.4,1.4,1)$.

\subsection{Transient behavior }

In transient states, 
the ordered  domains  with positive (negative) $\psi$  
tend to be elongated along 
the horizontal $x$ (vertical $y$) axis after their formation.  
They do not penetrate into the others   belonging to 
 the different variant (having the opposite $\psi$) 
on their encounters.   Furthermore, thickening 
of the stripe shape is prohibited 
by the fixed-volume condition $\av{e_1}=0$. 
As a result, our  system tended to a nearly pinned state 
at long times in  each simulation run.

In Fig. 10, we  display the time evolution of the total 
perimeter length in the relaxation from 
 disordered   to stripe intermediate states 
at the points C$_1$ (where 
$\tau/\tau_0=0.6$) 
and C$_3$(where 
$\tau/\tau_0=-1.4$) in 
the middle panel of Fig. 8. 
The initial stage of domain formation 
is in the range $t/t_0 \ls 10$ 
and the coarsening stops for $t/t_0 \gs 10^3$. 
At the higher  temperature (C$_1$),  
the ordered domains  consist of long stripes with a smaller 
areal fraction. 
At the lower temperature (C$_3$),  
the domains are distorted, consisting of both large and small 
ordered stripes.  
In Fig. 11,  we  display the relaxation  from  disordered  
 to lozenge  intermediate states 
at the points A$_1$ and B$_1$ in 
the left and middle panels  of Fig. 8. 
Here the domain pinning occurs at $t/t_0 
=2-4 \times 10^3$. The patterns are nearly 
symmetric at A$_1$ but 
largely asymmetric at B$_1$.  In these two cases,  the lozenge 
angle $\delta$ and the areal fraction $\phi$ 
are nearly the same (see Fig. 12), while the domain sizes 
are considerably different.

\subsection{Steady intermediate  states}

\begin{figure}
\begin{center}
\includegraphics[width=8cm]{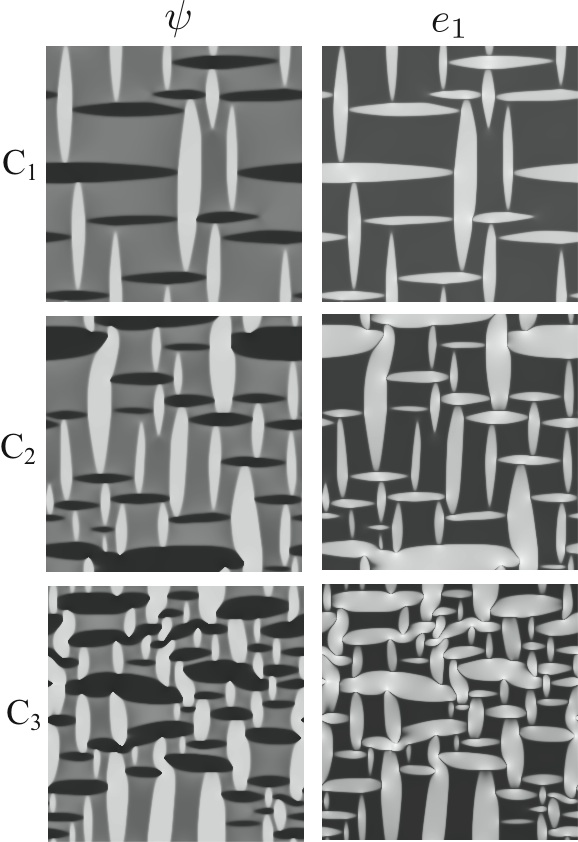}
\end{center}
\caption{Stripe patterns 
 of $\psi$ (left) and  $e_1$  (right) 
at the points C$_1$, C$_2$, and C$_3$, where 
$\tau/\tau_0=0.6$, $-0.4$, and $-1.4$, respectively, 
with  $\alpha/\alpha_0= 1.4$  and $v/v_0=1$ being common. 
In the left panels,  $\psi$ 
is positive  in the white regions, 
negative in the black  regions, 
and nearly zero  in the gray  regions. 
Our theory indicates $\psi=\psi_{\rm s},  -\psi_{\rm s}$, 
or  $0$ in the three regions.
 In the right  panels, 
$e_1$ is positive  in the white   regions 
and negative  in the black  regions. 
Theoretically, $e_1=e_1^0+e_{\rm s}  >0 $ 
or $e_1=e_1^0 <0$. 
See the corresponding $\phi$  
in the left panel of Fig. 12.}
\end{figure}

\begin{figure}
\begin{center}
\includegraphics[width=8cm]{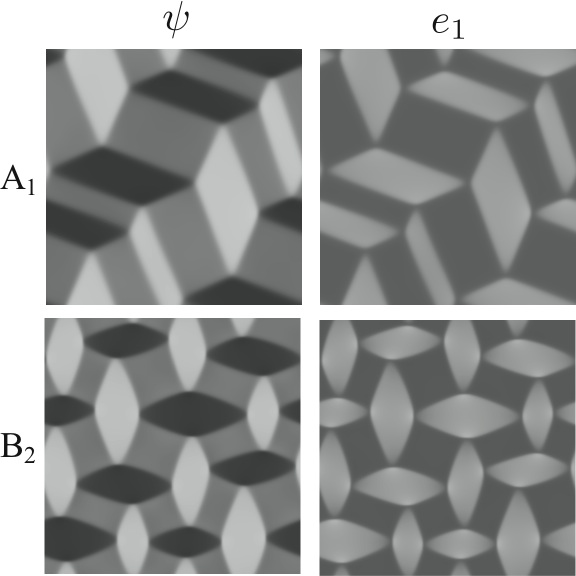}
\end{center}
\caption{ Lozenge patterns 
 of $\psi$ and  $e_1$ at the points  A$_1$ 
   and B$_2$ in  Fig. 8, 
where $(\tau/\tau_0,\alpha/\alpha_0,v/v_0)=
(0.6, 0.76,0.1)$  and 
($0.6,1,1)$, respectively. 
They are asymmetric ans symmetric, respectively, 
as in Fig. 5. 
At B$_2$ the edges are flattened, resulting in an increase 
of the ordered regions.}
\end{figure}

In our system,  the disordered phase ($\psi=0$)  is  
linearly stable for $\tau>g^2/\mu$ (as can be known from 
 Eq.(3.3)).  Therefore,      
we  lowered  $\tau$ below $g^2/\mu$              
to produce ordered domains.   
Notice that the intermediate phase 
can be stable or have a negative free energy 
for $\tau$ above  $g^2/\mu$ 
and  below the transition-point value, as can be seen in Figs. 8  
and 9.   To realize such intermediate states,  
we increased  $\tau$  starting  with twin  
or intermediate states. Then 
 $\phi$ decreased on approaching the transition point from below.

In Fig. 12, we show  $\phi$ 
 versus $(\tau-\tau_{\rm s})/\tau_{\rm sw}$ in  
the steady stripe states (left) and 
$\phi$  versus $(\tau-\tau_\ell)/\tau_{\ell {\rm w}}$ 
in the steady  lozenge states (right). 
Our theory predicts the linear 
dependence in  Eqs.(3.18) and (3.39).  
The points  A$_1$, A$_2$, B$_1$,  B$_2$, 
 B$_3$,  C$_1$, C$_2$, and C$_3$ 
correspond to those   in the  
phase diagrams in Fig. 8.  
Slightly below the transition, these   
 numerical data excellently agree 
with our theory.  
In addition, in the left panel,  
we  show  the  hysteresis behavior 
 dependent  on the initial condition in 
 the range $g^2/\mu<\tau<\tau_{\rm s}$. 
But  some discrepancies arise as $\tau$ approaches  
the lower bound of the temperature window. 
That is, as $\tau$ approach $\tau_{\rm s}-\tau_{\rm sw}$ 
in the left panel, 
$\phi$  become  smaller than predicted 
and tend  to saturate. 
In the right panel, 
 twin  states  were  
realized at  the points  A$_2$ and B$_3$, though they 
are  in the lozenge phase in our theory.

In Fig. 13, we show typical steady  patterns of 
$\psi$ and $e_1$ in the stripe   
states  C$_1$, C$_2$,  and C$_3$  
taken at  $t/t_0=5\times 10^3$.  
The patterns of  $e_2$ (not shown here) are 
almost indistinguishable from those of $\psi$.
The patterns in Fig. 13 consist of long stripes for small $\phi$,  
but they contain  even 
short   ones and are considerably 
distorted for large $\phi$.  
Thus, as  $\tau$ approaches 
$\tau_{\rm s}-\tau_{\rm sw}$, the domain morphology  
  becomes  more complex than 
in our theoretical picture.  This  explains why 
the data of $\phi$ in the left panel 
of Fig. 12 are  below  the theoretical straight line.

In Fig. 14,  the lozenge pattern 
at   the point A$_1$  
is  asymmetric  with rectangular  disordered regions, 
while that at  the point B$_2$ 
is more symmetric  and  some disordered regions 
 are  nearly square.  The  $\phi$ for the pattern at B$_2$ 
is $0.51$ and is larger  than those at A$_1$ and B$_1$ 
  in Fig. 12 (being 0.42 and 0.43, respectively)  by 20$\%$.  
The shapes of the lozenges  at  B$_2$ are more 
rounded than those at  A$_1$  and B$_1$ 
with flattened edges. 
As a result, $\phi$ can be  larger at 
 B$_2$ than at B$_1$.

We stress that the intermediate patterns strongly 
depend on  how they were created. They are 
 sensitive even to the initial noise. 
For example, Figs. 11 and 14 
show the   lozenge 
patterns  at the same point A$_1$ at $t/t_0=5\times 10^3$, 
where the large difference arises only from the initial noise. 
However, the pattern in Fig. 14 
for A$_1$ will change into a more symmetric one 
at longer times. Note that  the lozenge angle $\delta$ 
and the areal fraction $\phi$ are rather insensitive to 
the history. 
In fact, $\phi=0.41$ and $0.42$ 
for the point A$_1$ in Figs. 11 and 14, respectively.

\section{Summary and remarks} 

Though in 2D, we have presented a theory 
of intermediate states at  structural 
phase transition using  a minimal model 
 with  a one-component order parameter $\psi$ twofold 
coupled to the  tetragonal strain properly $(g \neq 0)$  and 
to the dilational strain improperly ($\alpha \neq 0)$. 
The pinning of the domains is due to the nonlinearity 
(third and fourth order terms in the free energy) 
and there is no impurity.  
We summarize our main results.\\
(i) In Eq.(2.27) or in Eq.(3.3) we have derived the elastic  
free energy contribution  $F_{\rm e}$  expressed 
in terms of $\psi$. It is very complicated due to the simultaneous 
presence of the proper coupling $(\propto g)$ 
and  the improper coupling $(\propto \alpha)$.\\ 
(ii) For the stripe intermediate phase,  the order 
parameter  $\psi=\psi_{\rm s}$ 
within the ordered domains is determined  
by Eq.(3.15).  The areal fraction 
$\phi$, the transition reduced temperature $\tau_{\rm s}$, 
and the width of the intermediate 
region $\tau_{\rm sw}$ are expressed as in Eqs.(3.18)-(3.20). 
The free energy decrease is of the simple form Eq.(3.22).\\
(iii)  For the lozenge intermediate phase,  the order parameter 
 $\psi=\psi_\ell$ 
within the ordered domains is given by  Eq.(3.38). 
 The strains are approximately given by  Eq.(3.35), where 
  $\Gamma$ in Eq.(3.30) is  small within  the 
ordered domains. The areal fraction 
$\phi$, the transition reduced temperature $\tau_\ell$, 
and the width of the intermediate 
region $\tau_{\ell {\rm w}}$ are given  in Eqs.(3.39)-(3.41). 
The free energy decrease is of the simple form Eq.(3.43).\\
(iv) We have compared the free energies in 
Eqs.(3.22), (3.43), (3.48), 
and (3.52) for the intermediate and ordered phases 
in the plane of $\tau$ and $\alpha$ with the other parameters held fixed.  
The transition reduced temperature, 
the phase diagrams,    and the bird views 
 of $\psi$ and $\phi$ are in Figs. 7-9, respectively.  
The intermediate states can appear 
in the case  $\alpha \gs \alpha_0$, where $\alpha_0$ 
is given in Eq.(2.30) 
and can be small near the tricritical point. 
Note that the ratio of the 
dilational  strain and the tetragonal 
strains is of order $\mu/K$ for $\alpha\sim \alpha_0$.  
\\ 
(v) We  have presented simulation results in 2D by 
integrating the dynamic equation (2.12). 
Figures  10 and 11  illustrate  
the freezing processes 
of the domain growth resulting 
in mesoscopic intermediate states. 
Figure 12 displays the areal fraction of the ordered regions 
obtained numerically. Figures 13 and 14 give typical 
 patterns of the stripe and lozenge intermediate phases. 
These simulation results are 
consistent with our  theory.\\

Further we give some remarks.\\
(1) In calculating the free energy $F$ in the  
intermediate states, we have neglected the surface tension 
contribution. This should be justified when 
the domain size much exceeds the surface thickness 
of order $\ell$ in Eq.(2.29).\\ 
(2) The previous 2D  simulation 
in the strain-only theory \cite{Onukibook,pre} 
already produced the lozenge patterns in the presence of the third 
order term proportional to $e_1e_2^2$, which plays 
the same role as the dilational 
improper coupling in this paper. 
\\
(3) If a solid  is  compressed  along one of the principal axes, 
the tetragonal extension along the compression 
axis  becomes  energetically  unfavorable \cite{Onukibook,JT}, 
leading to    a square-to-rectangle transition perpendicular 
to the axis. Then the  transition  could be described by 
a 2D  theory with  a one-component order parameter, 
although the lateral elastic deformations 
are much more complicated in real epitaxial films 
 than treated in this paper \cite{Desai}.\\
(4) In its present form,  
our theory cannot explain 
the 3D experiments  
\cite{Na,Ohshima,Shapiro,Seto,Kinoshita,Uezu,Waseda}. 
We should construct  a 3D theory of intermediate states.  
We will shortly report  simulation results 
of intermediate states in 3D.\\

\section*{Acknowledgments}
This work was supported by Grants in Aid for Scientific Research and for
the 21st Century COE project (Center for Diversity and Universality in
Physics) from the Ministry of Education, Culture, Sports, Science and 
Technology of Japan.

\appendix 
\section{Order parameter in stripe intermediate states}
\setcounter{equation}{0}
\renewcommand{\theequation}{A.\arabic{equation}}

We here solve the cubic equation (3.15). 
For $u>0$ and $v=0$, we  simply 
 have  $\psi_{\rm s}=2\alpha g/uL$.    
Including the case of $u<0$,  we  may generally solve it 
 in the scaling form, 
\be 
\psi_{\rm s}= \psi_{\rm t} \Phi(u/u_{\rm t}),
\en 
where we define two quantities, 
\bea 
\psi_{\rm t}&=& (3\alpha g/2Lv)^{1/3},\\
u_{\rm t} &=& 4v\psi_{\rm t}^2/3= (v/3)^{1/3}  (4\alpha g/L)^{2/3}.
\ena 
The scaling function $\Phi(w)$ is  determined by  
\be 
\Phi(w)^3+w\Phi(w)=1
\en 
with $\Phi(w)>0$ (since $\psi_{\rm s}>0$). This cubic equation 
can be solved to give an explicit expression for $\Phi(w)$. 
For $w>-3/4^{1/3}\cong -1.89 $,  it reads   
\be 
\Phi=  \bigg(\sqrt{\frac{w^3}{27}+\frac{1}{4}}+\frac{1}{2}\bigg)^{1/3}- 
 \bigg(\sqrt{\frac{w^3}{27}+\frac{1}{4}}-\frac{1}{2}\bigg)^{1/3}. 
\en  
For $w<-3/4^{1/3}$,  it is expressed as 
\be 
\Phi=  2\bigg|\frac{w}{3}\bigg|^{1/2} \cos 
\bigg [\frac{1}{3}\cos^{-1}\bigg(\frac{1}{2}\bigg|\frac{3}{w}
\bigg|^{3/2}\bigg)\bigg] . 
\en 
We   find $\Phi(w) \cong 1/w$   for $w \gg 1$  and 
$\Phi(w) \cong |w|^{1/2}$ for $w \ll -1$, so that 
\bea 
\psi_{\rm s}&\cong& 2\alpha g/uL \quad (u \gg u_{\rm t}),\nonumber\\
&\cong &   (3|u|/4v)^{1/2} \quad ((u \ll -u_{\rm t}).
\ena 
The first line holds for $ {\bar u}\gg u_{\rm t}+\beta_L$, 
while the second line  for $\beta_L\gg {\bar u}+ u_{\rm t}$.

\end{document}